\documentclass[twocolumn,showpacs,showkeys,amsmath,amssymb]{revtex4}

\usepackage{graphicx}
\usepackage{dcolumn}
\usepackage{bm}

\begin{document}

\title{Quantitative calculations of charge carrier densities in the depletion layers at
YBa$_2$Cu$_3$O$_{7-\delta}$ interfaces}

\author{U.~Schwingenschl\"ogl}
\author{C.~Schuster}
\affiliation{Institut f\"ur Physik, Universit\"at Augsburg, 86135 Augsburg,
Germany}

\date{\today}

\begin{abstract}
Charge redistribution at high-T$_c$ superconductor interfaces and grain boundaries on the one
hand is problematic for technological application. On the other hand, it gives rise to a great
perspective for tailoring the local electronic states. For prototypical (metallic) interfaces,
we derive {\it quantitative} results for the intrinsic doping of the CuO$_2$-planes, i.e.\
for the deviation of the charge carrier density from the bulk value. Our data are based on
ab-initio supercell calculations within density functional theory. A remarkable hole-underdoping
is inherent to the clean interface, almost independent of the interface geometry. On the
contrary, cation substitution as well as incorporation of electronegative impurities can
compensate the intrinsic charge transfer and provide access to an exact adjustment of the
superconductor's doping. The effects of oxygen deficiency are discussed.
\end{abstract}

\pacs{73.20.-r,73.40.Jn,74.25.Jb,85.25.Am}
\keywords{density functional theory, interface, grain boundary, intrinsic doping, charge transfer}

\maketitle

In recent years, interfaces incorporating a high-T$_c$ superconducting material are
subject to a quickly growing interest, in both the experimental and theoretical community. Due
to the possibility of orbital reconstructions and a modified chemical bonding at interfaces, design of
heterostructures with engineered physical properties has become a prosperous field. Much
effort has focussed on YBa$_2$Cu$_3$O$_{7-\delta}$ based interfaces, often attached to
oxide compounds \cite{chakhalian07,carrington07,elfimov08}. In the high-T$_c$ cuprates and
nickelates, even a very small change of the crystal structure or the chemical composition, or
the presence of structural strain or disorder, can result in fundamental alterations of the
electronic properties \cite{mannhart98,hilgenkamp02,lang02,stripes}. From the structural
point of view, large dielectric constants and small carrier densities amplify relaxation
effects \cite{samara90}. This fact is of special importance for the transport properties of the
superconductor, since the local charge distribution in the CuO$_2$-planes is seriously
influenced near interfaces. Moreover, as a consequence of electrostatic screening lengths of a few
nanometers and a strong structural inhomogeneity, conventional band bending models cannot
be applied for describing the charge redistribution.

Both the structural relaxation and the local electronic states at interfaces of high-T$_c$ compounds
have been investigated for specific situations by various methods, including superlattices composed
of different high-T$_c$ compounds \cite{varela99}, doping induced by charge transfer from the localized
states in a hopping insulator \cite{kozub06}, and cuprates in contact to manganites \cite{yunoki07}.
In this context, we have recently dealt with the influence of the charge redistribution at
a metal interface on the doping in YBa$_2$Cu$_3$O$_7$ (YBCO).
It turnes out that the superconducting CuO$_2$-planes are subject to a strong
intrinsic electron-doping, where screening effects are important for describing the charge transfer
amplitude \cite{us07epl}. The net charge transfer amounts to 0.13 electrons per Cu site when
the interface is oriented parallel to the CuO$_2$-planes, and to 0.09 electrons for a perpendicular
orientation \cite{us07apl}. In addition to this weak dependence on the geometry, the charge
transfer is also robust against the specific metal forming the interface, thus giving rise
to a ``universal'' behaviour. In this paper we deal with the question, how the intrinsic
doping changes under different kinds of perturbations of the interface. In particular, we give a {\it quantitative}
evaluation of various mechnisms that can be used for {\it tailoring} the charge transfer, which
provides a great technological potential for material optimization and design.
\begin{figure}
\begin{center}
\includegraphics[width=0.48\textwidth,clip]{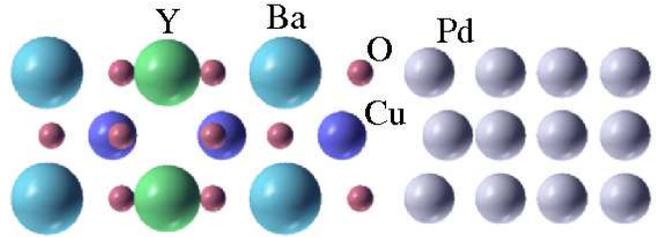}
\end{center}
\caption{\label{fig-ybco} (Color online) Interface between bulk Pd and YBCO (precisely: the CuO-chain layer).
The CuO$_2$-planes run parallel to the interface.}
\end{figure}

While for bulk high-T$_c$ compounds a formidable number of investigations of impurity effects
is found in the literature, similar results for surfaces and interfaces are rather rare. In
addition, studies of interfaces have predominately addressed the domain of the superconductor, whereas
little deal with the interface itself. On the contrary, very much attention has focussed on the influence
of the O content -- in the full range of the phase diagram \cite{shibata00}.
Moreover, it has been observed that the superconducting properties of YBCO strongly alter
under substitution of Y by Ca. The critical current at low magnetic fields and the critical temperature 
decrease with the Ca concentration \cite{slyk}. In the case of grain boundaries, Ca doping
hence is used to compensate the local charge redistribution \cite{schmehl99}.
The YBCO crystal structure is subject to drastic alterations under the incorporation of defects.
Especially, the lattice parameters change considerably as a function of the O stochiometry \cite{kogachi88}.
Experiments in the late eighties had already indicated that the insertion of F/Cl into underdoped
YBCO can enhance the critical temperature \cite{ovshinsky87}, where the F/Cl atoms tend to occupy
the empty metal sites in the CuO-chains \cite{graffe88,amitin93}. Interestingly, this insertion of
halogenides is accompanied by a higher hole content in the CuO$_2$-planes, which suggests that a
F/Cl impurity has the potential to re-extract the excess charge arising at an YBCO interface.

Our data are obtained from density functional theory, applying the generalized
gradient approximation and the Perdew-Burke-Ernzernhof scheme.
We use the WIEN$2k$ program package \cite{wien2k}, with a mixed linearized augmented
plane wave and augmented plane wave plus local orbital basis. Moreover, we set $RK_{\rm max}=7$
when we investigate Cl impurities and $RK_{\rm max}=4.7$ otherwise. The radii of the muffin tin
spheres are (in Bohr radii): 1.8 for Cu, 2.2 for Cl, 2.25 for Ca, 2.3 for Y, 2.5
for Ba, 1.45 to 1.55 for O, and 2.1 to 2.4 for Pd.

Because band-bending takes place on the length scale of
the YBCO $c$ lattice constant, the electronic structure of an interface is
accessible to a supercell approach with periodic boundary conditions \cite{supercell}.
It is convenient to use as the metallic substituent fcc Pd, which has a minimal lattice
mismatch of only 0.7\% with respect to the YBCO $a$/$b$ lattice constant. We have
checked that our results do not depend on this choice by comparing the charge
transfer at YBCO-Pd and YBCO-Ag interfaces, for which we find almost identical data in
all relevant respects \cite{us07epl,us07apl}. For illustrating the principle of
the supercell setup, Fig.\ \ref{fig-ybco} displays the near-interface region of the supercell for
a clean YBCO-Pd interface oriented parallel to the CuO$_2$-planes. It comprises two YBCO unit cells
terminated by the CuO-chain layer, as indicated by experiment \cite{xin89,derro02}, and four Pd fcc
unit cells stacked in [001]-direction. The supercells under consideration all have a similar size.

\begin{figure}
\begin{center}
\includegraphics[width=0.37\textwidth,clip]{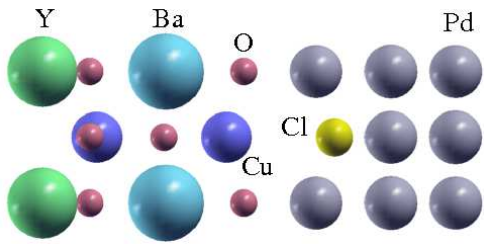}
\end{center}
\caption{\label{fig-struct} (Color online) Cl doped YBCO interface. Every 2nd contact metal atom is substituted by
Cl. The orientation of the CuO$_2$-planes is parallel to the interface, which is formed by CuO-chains.}
\end{figure}
We use the experimental YBCO bulk lattice constants
$a=3.865$\,\AA\ and $b=3.879$\,\AA\ \cite{siegrist87}, which is likewise applied to the Pd region in
order to avoid lattice mismatch in the contact plane. In the perpendicular direction,
the standard Pd lattice constant $c=3.89$\,\AA\ is an adequate choice. By the minor dependence
of the charge transfer on the interface geometry, all subsequent results refer to the setup
of Fig.\ \ref{fig-ybco}. In the case of fully oxydized YBCO, one half of the topmost
Pd atoms consequently has an O bonding partner in the interface CuO-chain layer, while the
other half is located next to an unoccupied site. We count atomic layers with respect
to the interface, i.e.\ the first Pd layer is attached to the first atomic layer of
the YBCO cell, formed by the the CuO-chains.
\begin{table}
\begin{tabular}{l|c|c|c}
& \hspace{0.09cm}YBCO bulk\hspace{0.09cm} & \hspace{0.09cm}YBCO interf\hspace{0.09cm}
& \hspace{0.09cm}YBCO$_{6.75}$ interf\hspace{0.09cm}\\\hline
d$_{\rm Cu-Pd}$    & -- &3.32 & 2.83\\
d$_{\rm O-Pd/Cl}$  & -- &2.12, 3.87 & 2.14\\
d$_{\rm Cu-O_{interf}}$     & 1.94 & 2.02 & 1.88\\
d$_{\rm Cu-O_{Ba}}$& 1.90 & 1.90 & 1.87 
\end{tabular}\\[0.5cm]
\begin{tabular}{l|c|c}
& Y$_{0.75}$Ca$_{0.25}$BCO interf & Cl doped YBCO interf\\\hline
d$_{\rm Cu-Pd}$    & 2.76 & 3.28\\
d$_{\rm O-Pd/Cl}$  & 1.98 & 2.12, 3.83\\
d$_{\rm Cu-O_{interf}}$     & 1.94 & 2.01\\
d$_{\rm Cu-O_{Ba}}$& 1.86 & 1.84 
\end{tabular}
\caption{Selected bond lengths (in \AA) for different interfaces (see the text for details),
obtained by structure optimization. Bulk YBCO bond lengths are given for comparison.}
\label{tab1}
\end{table}

The following 3 mechanisms of tailoring
the intrinsic doping are investigated: First, the incorporation of electronegative ions is realized by substituting
each second interface Pd atom without an O neighbour by a Cl atom. This arrangement is shown in
Fig.\ \ref{fig-struct}, after structure optimization. Second, cation substitution is studied for the replacement
of Y by Ca, where we choose a Ca content close to the optimal value of $\approx$ 30\% for grain boundary doping
\cite{schmehl99}. To keep the computational effort manageable, we replace
each fourth Y atom by Ca, i.e.\ we deal with Y$_{0.75}$Ca$_{0.25}$BCO.
Third, the effects of O deficiency are addressed for YBCO$_{6.75}$, where each
fourth interface O atom, all from CuO-chains, is removed. We note that the system therefore stays
far in the metallic regime \cite{dagotto94}, so that valid results are expected from a band
structure calculation using the present approximations.

Turning to the chemical bonding in the vicinity of the interface, there is always the same, expected behaviour. We find
a strong tendency towards Pd--O bonding and a Cu--Pd repulsion. The Cu--Cl repulsion is weaker.
Moreover, the Cl incorporation induces only weak structural distortions of the superconductor,
whereas the first two Pd layers are subject to a finite relaxation. In fact, the relaxation in the
superconductor domain declines quite quickly for all
interfaces under investigation, so that the second CuO$_2$-plane
is virtually undistorted. This is particularly true for the O deficient case where we may anticipate
a different behaviour. Selected bond lengths are compared to each other in Table \ref{tab1}.
Reflecting covalent Pd--O bonding across the interface, the nearby Pd sites are subject to shifts of the
valence electronic states from lower energy to about $-2$\,eV, with respect to the Fermi energy.
This gives rise to a pronounced structure in the Pd $4d$ density of states (DOS), characterized
by a significant Pd--O hybridization. However, even
for the second Pd layer, the DOS resembles the bulk Pd $4d$ DOS very well, and a nearly
perfect agreement is obtained for the third layer. As a consequence, structural distortions are
screened quickly on both sides of the interface, which a posteriori justifies the size of our supercells.

\begin{figure}[t]
\begin{center}
\includegraphics[width=0.39\textwidth,clip]{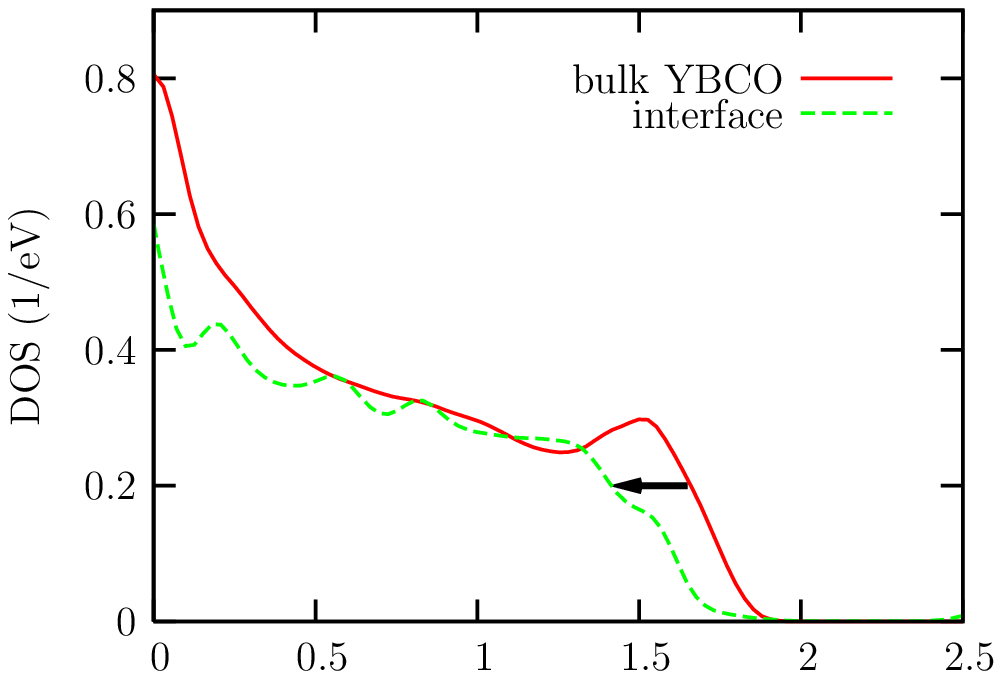}\\
\includegraphics[width=0.39\textwidth,clip]{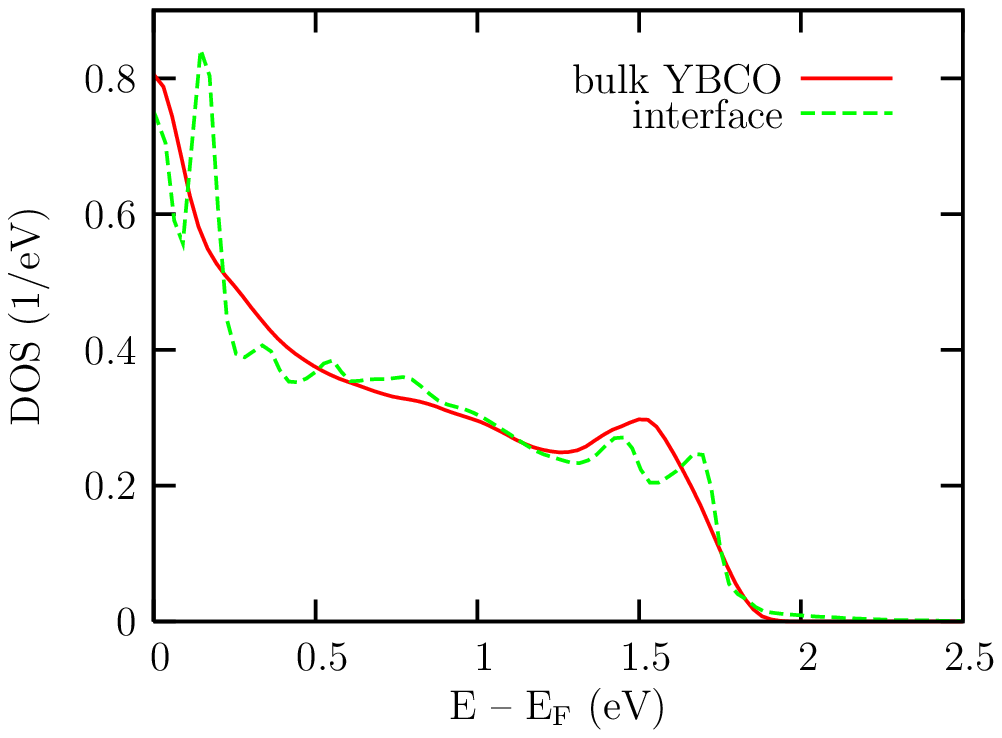}
\end{center}
\caption{\label{fig-Cl} (Color online) Partial Cu $3d$ DOS for the 2nd CuO$_2$-plane off the
clean (upper panel) and Cl doped (lower panel) YBCO interface, as
compared to the bulk DOS.}
\end{figure}
\begin{figure}[t]
\begin{center}
\includegraphics[width=0.39\textwidth,clip]{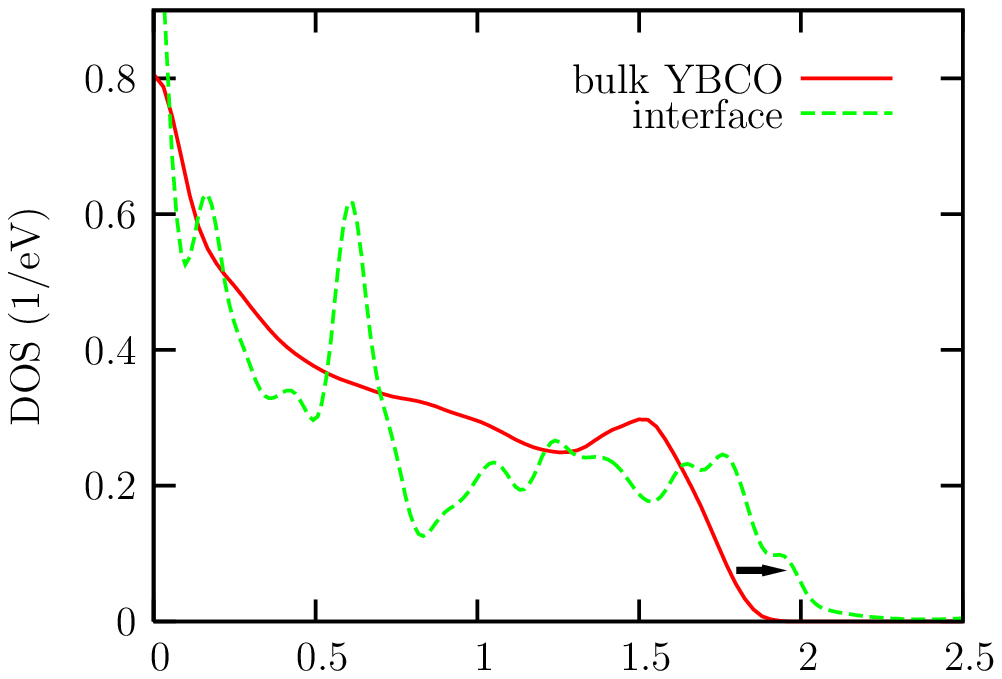}\\
\includegraphics[width=0.39\textwidth,clip]{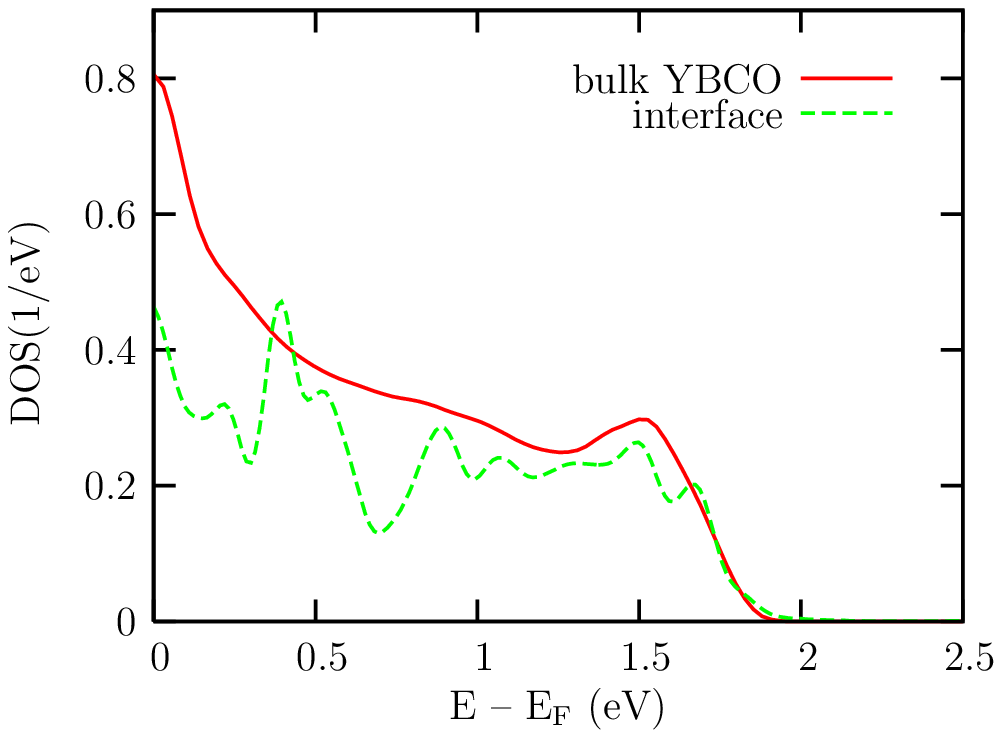}
\end{center}
\caption{\label{fig-O} (Color online) Partial Cu $3d$ DOS for the 2nd CuO$_2$-plane off the
Y$_{0.75}$Ca$_{0.25}$BCO (upper panel) and O-deficient YBCO$_{6.75}$ (lower panel) interface.
In each case, the findings are compared to the corresponding DOS of bulk YBCO.}
\end{figure}

We first investigate the effect of the Cl doping on the YBCO electronic states.
Cl--Pd hybridization is tiny for atoms in the first Pd layer, according to
an ionic bonding \cite{ions1,ions2}. However, there are remarkable admixtures of the
nearest neighbour Pd atom (see Fig.\ \ref{fig-struct}) in the Cl dominated DOS range, which
we attribute to enhanced bonding because of a very short distance. Of course, it is to be expected that the Cl extracts
charge from the surrounding area until a Cl$^-$ ion state is reached.
Since no reduction of the Pd charge is observed under Cl doping, the extraction
should affect only the YBCO region. Assuming that the charge reduction affects all 7 Cu atoms per unit cell to a
similar degree, naive electron counting indicates that the intrinsic charge transfer due to
the interface (0.13 electrons per Cu site) can be compensated by this mechanism quite well.

We therefore compare in Fig.\ \ref{fig-Cl} the DOS for the clean and for the Cl doped interface to bulk YBCO.
The data refer to the second CuO$_2$-plane, because here the structural relaxation has already decayed.
According to Fig.\ \ref{fig-Cl}, the Cu $3d$ DOS confirms our speculation. Whereas
a strong reduction of the (unoccupied) states is visible for the clean interface, this charge
transfer is fully compensated for the Cl doped case. The areas under the DOS curves in
the lower panel of Fig.\ \ref{fig-Cl} coincide almost perfectly, reflecting identically doped
CuO$_2$-planes. In fact, the suppression of charge transfer to the YBCO domain due to
the Cl atoms is surprisingly strong. Electronegative impurities consequently seem to form a
rather rigid potential barrier which prohibits any intrinsic doping.

Another way for overcoming the hole-underdoping at interfaces is simultaneous overdoping:
replacing Y$^{3+}$ by Ca$^{2+}$. The initial
doping state should be restored, since the introduction of Ca increases the hole concentration.
The upper panel of Fig.\ \ref{fig-O} confronts the DOS of the pure YBCO interface with
the results for an Y$_{0.75}$Ca$_{0.25}$BCO interface. Whereas differences in the
Cu $3d$ DOS shape reflect remarkable alterations of the electronic states, in
particular of the Cu--O bonding, their occupation fulfills exactly the trends mentioned
before. The interface doping is even overcompensated for the configuration under investigation,
which is visible in Fig.\ \ref{fig-O} by a shift of the DOS curve to higher energies.
Further evaluation of the data gives a charge reduction of 0.08 electrons. 25\% Ca doping
thus comes along with an injection of 0.21 additional holes per Cu site into the CuO$_2$-planes.

The fact that local Ca doping of grain boundaries can enhance the supercurrent
density has been attributed to a reduction of the charge accumulation at grain boundary dislocations,
which causes a hole depletion \cite{schofield04}. The size
of the perturbed regions seems to be smaller for Ca doped samples and the interface conductivity
thus is improved. Resistivity measurements in fact reveal changes of the height and the shape
of the electrostatic potential barrier under Ca doping \cite{mennema05}. However,
the experiments likewise indicate that such modifications cannot explain the observed
effect of Ca doping on the transport properties. It therefore has been speculated that extra holes must
be generated for some reason, which finally is substantiated by our findings. A grain boundary
appears to be quite well described in terms of an interface between an unperturbed superconductor domain and
the charge reservoir of the grain boundary dislocations. Ca doping then compensates
the intrinsic charge transfer present at this interface and so introduces additional holes.

As is known from many experiments, O deficiency has a very strong
influence on the hole content of the CuO$_2$-planes, see \cite{temmerman01},
for instance, and the references therein. In analogy to our previous
proceeding, we now quantify this effect. Evaluating the DOS given in
the lower panel of Fig.\ \ref{fig-O}, we obtain a reduction of the
hole count in our YBCO$_{6.75}$ interface by 0.10 holes per Cu
site. However, this reduction is not due to charge transfer,
as we do not observe an energetical shift of the DOS curve, but
traces back to a modified Cu--O bonding.

We have discussed electronic structure calculations for prototypical interfaces
of a high-T$_c$ superconductor and a normal metal. In each case, we have
evaluated {\it quantitatively} the interface charge transfer from the metal into the
YBa$_2$Cu$_3$O$_{7-\delta}$ domain. The intrinsic underdoping of 0.13 holes per Cu atom
(for $\delta=0$), is fully suppressed by electronegative impurities. Cation substitution
in the superconductor likewise leads to a (over-)compensation, where the initial hole
concentration is restored for a 15\% Ca content. Since O deficiency at the
interface has just the opposite effect, it is therefore possible to exactly design the
doping state of the superconductor. Reduction of the O content by $\delta=0.25$ leads to
an underdoping of 0.10 holes per Cu atom. We expect that all these values likewise
apply to the intentional doping of grain boundaries, which paves the way for a systematic
optimization of transport properties.

\subsection*{Acknowledgement}
We acknowledge valuable discussions with U.\ Eckern, V.\ Eyert, J.\ Mannhart, and T.\ Kopp.
Financial support of the work was provided by the Deutsche Forschungsge\-meinschaft (SFB 484).

\end{document}